\documentclass[aps,prl,twocolumn,showpacs,superscriptaddress,groupedaddress]{revtex4-2}
\pdfoutput=1

\usepackage[utf8]{inputenc}
\usepackage[T1]{fontenc}
\usepackage{xcolor}
\usepackage{amssymb}
\usepackage{amsfonts}
\usepackage{booktabs}
\usepackage{enumerate}
\usepackage[usenames,dvipsnames]{xcolor}  %%%%%%% usenames,dvipsnames required to get BrickRed
\usepackage{colortbl} 
\usepackage{graphicx} %%%%%% do not use \usepackage[dvips]{graphicx}%%%% it will creat a clash with color n brickred will not appear%%%%%
\definecolor{lightblue}{rgb}{0.93, 0.95, 1.0} % Azul claro
\definecolor{lightgray}{gray}{0.9} % Gris claro
\usepackage{bbold}
\usepackage{amsthm}
\usepackage{amsmath}
\usepackage{slashed}
\usepackage{amssymb}
\usepackage{cancel}
\usepackage{multirow}%\usepackage{dsfont}
\usepackage{hhline}
\usepackage{float}
\usepackage[utf8]{inputenc}
\usepackage{orcidlink}
\usepackage{booktabs}
\usepackage{ulem}
\definecolor{darkred}{rgb}{0.6,0,0}
\usepackage[]{hyperref}

\usepackage{braket}
\definecolor{linkcolor}{rgb}{0,0,0.5}
\allowdisplaybreaks
\allowdisplaybreaks[1]
\allowdisplaybreaks[2]

\usepackage{soul}
\usepackage{url}

\definecolor{darkviolet}{rgb}{0.58, 0.0, 0.83}

\definecolor{orange}{rgb}{1,0.5,0}

\usepackage{soul}

 \newcommand{\AddrCFTP}{%
  Instituto Superior T\'{e}cnico, Universidade de Lisboa, 
		\small Av. Rovisco Pais 1, 1049-001 Lisboa, Portugal\vspace{0.1cm}}

\newcommand{\AddrBasel}{Department of Physics, University of Basel, Klingelbergstrasse 82, CH-4056 Basel,Switzerland}

\begin{document}

\title{\boldmath \color{BrickRed} Gravity tidings from domain walls: \\
Flavour hierarchies are making waves}
\author{Stefan Antusch\orcidlink{0000-0001-6120-9279}}\email{stefan.antusch@unibas.ch} \affiliation{\AddrBasel}
\author{Ivo~de~Medeiros~Varzielas\orcidlink{0000-0003-4996-3484}}\email{ivo.de@udo.edu}
\affiliation{\AddrCFTP}
\author{Miguel Levy\orcidlink{0000-0002-5854-8939}}\email{miguelpissarra.levy@unibas.ch}
\affiliation{\AddrBasel}

%%%%%%%%%%%%%%%%%%%%%%%%%%%%%

\begin{abstract}

Explaining the observed charged fermion mass hierarchies points to flavour symmetries inducing a suppression of the lighter species' masses. When the symmetries are global, it is expected that such symmetries are broken by gravity via Planck scale suppressed effective operators. The potential of the spontaneous symmetry-breaking ``flavon’’ field, if the symmetry is discrete, then possesses several minima, with the vacuum-degeneracy lifted by the gravity effects. In such scenarios, domain walls might be generated in the process of symmetry breaking. Due to the bias, however, they potentially annihilate sufficiently before Big Bang nucleosynthesis, avoiding conflict with observations and generating a characteristic contribution to the stochastic gravitational wave background. We discuss whether and how minimalistic supersymmetric and non-supersymmetric realisations of such theories can give rise to observable gravitational waves. 

\end{abstract}

%\tableofcontents

%%%%%%%%%%%%%%%%%%%%%%%%%%%%%

\maketitle

%%%%%%%%%%%%%%%%%%%%%%%%%%%%%%%

%%%%%%%%%%%%%%%%%%%%%%%%%%%%%%%%%%%%%%%%%%%%%%
\section{Introduction}
\label{section: Introduction}
%%%%%%%%%%%%%%%%%%%%%%%%%%%%%%%%%%%%%%%%%%%%%%

Particle physics is undoubtedly shaped by symmetries.  
Discrete symmetries in particular have taken a central role in explaining open questions of the flavour sector.  
The inability of the Standard Model (SM) to provide a fundamental reason for the number of generations, mixing patterns, and sizeable hierarchies across the generations of fundamental fermions is referred to as the Flavour Puzzle. The most promising solution is arguably the extension of the SM with flavour symmetries. One of the simplest possibilities is relying on the Froggatt-Nielsen (FN) mechanism~\cite{Froggatt:1978nt}, where, in its original formulation, a $U(1)$ flavour symmetry accounts for the hierarchies of the quark sector.  
Adapting the FN mechanism to discrete
symmetries is straightforward.

With discrete symmetries, the vacuum structure will feature degenerate minima, that
combined with the evolution of the early Universe, featuring e.g.\ a thermally induced phase transition or a phase transition ending inflation, can give rise to topological defects such as cosmic strings (CSs), and domain walls (DWs).  
The latter 
arise if the theory has sets of minima related by discrete (global) symmetries.  
If the symmetries are somehow broken, the degeneracy of the minima is no longer necessarily present. Patches of the Universe in energetically disfavoured local minima are bounded by unstable DWs.
The annihilation of DWs releases energy, giving rise to gravitational waves (GWs) which can be observable in experiments (see e.g.\ ~\cite{Hiramatsu:2012sc, Hiramatsu:2013qaa, Kawasaki:2014sqa, Wu:2022stu, Wu:2022tpe, Jueid:2023cgp, King:2024lki, Fu:2024rsm, Fu:2024jhu,  Gouttenoire:2025ofv, Li:2025gld,  Notari:2025kqq, Borah:2025bfa, Fu:2025qhf, Barreto:2026igt} and e.g. \cite{Saikawa:2017hiv} for a review).

The goal of this work is to study the GW signals that can arise from $Z_N$ symmetries, and link those general results to the particular case of the simplest discrete FN (dFN) models.

%%%%%%%%%%%%%%%%%%%%%%%%%%%%%%%%%%%%%%%%%%%%%%
\section{A Minimal Discrete Froggatt-Nielsen}\label{section: dFN}
%Mechanism}
%%%%%%%%%%%%%%%%%%%%%%%%%%%%%%%%%%%%%%%%%%%%%%

Froggatt-Nielsen models rely on Yukawa sectors arising from the spontaneous symmetry breaking of a flavour symmetry, explaining the SM fermion hierarchies without hierarchical dimensionless couplings.  
Their simplest realisation employs a single FN-symmetry breaking field $\chi$ (a flavon), and the effective couplings arise as an $\mathcal{O}(1)$ coupling suppressed by $\varepsilon^n$, where $\varepsilon = \langle \chi \rangle/\Lambda$ (with $\Lambda$ the messenger scale) and $n$ is the order of the effective operator in flavon insertions. 
Since $n$ must be a (positive) integer, $\varepsilon$ is linked to the smallest hierarchy we wish to explain with the FN mechanism.  
In most cases, this is the Cabibbo angle: $\lambda_C \approx 0.225$~\cite{ParticleDataGroup:2024cfk}.  
Taking $\varepsilon = \lambda_C$, the first generation of fermions require the effective operators to have $n \sim 8$.   

In standard FN models, the $n$ for the leading order contribution is adjusted by the $U(1)$ charges of the fields. 
This is not the case for dFNs, as we must require $\bar{\psi}_1 \chi^n H \psi_2$ remains the dominant contribution, 
with both $\bar{\psi}_1 \chi^{m<n} H \psi_2$ and $\bar{\psi}_1 (\chi^\dagger)^{m<n} H \psi_2$ not invariant.  
The second condition is trivially satisfied in Supersymmetry (SUSY) due to the requirement of holomorphicity of the superpotential.  
The first condition selects the $G_f = Z_N$, with $N \geq n+1$ (we assume that $\chi$ breaks $Z_N$ completely, such that we can always choose $\chi \sim \omega$, with $\omega^N=1$).  
If we do not wish to impose SUSY, combining both conditions leads to $G_f = Z_N$, with $N \geq 2n$, because 
without the protection of holomorphicity, $\bar{\psi} H \psi \chi^n$, with $n>N/2$ is subleading compared to $\bar{\psi} H \psi (\chi^\dagger)^{N-n}$.  
Insisting on having $n=8$ suppressions, SUSY minimally requires $G_f = Z_9$, whereas non-SUSY models would have $G_f = Z_{16}$ (smaller $N$ can still be relevant for dFN models with multiple $Z_N$-breaking fields). 

From an aesthetic point-of-view, and to have the theory be sturdy against other sources of $G_f$ breaking (either soft or explicit), smaller $Z_N$ symmetries are preferable.  
A workaround is to have $\lambda_C \approx 2 \mathcal{O}(1) \varepsilon$ (as the mixing angles can receive contributions from both sectors), with $\varepsilon \sim \lambda_C^2$.  

This has been shown to work, and so we focus on the texture of refs.~\cite{Antusch:2023shi, Greljo:2024evt}: 
\begin{equation}\label{eq:FNstructure}
    Y_u \sim 
    \begin{pmatrix}
        \varepsilon^4   & \varepsilon^3 & \varepsilon^2 \\ 
        \varepsilon^3   & \varepsilon^2 & \varepsilon \\ 
        \varepsilon^2   & \varepsilon   & 1
    \end{pmatrix} \, , 
    \quad 
    Y_d \sim \varepsilon^2
    \begin{pmatrix}
        \varepsilon^2   & \varepsilon^2 & \varepsilon^2 \\ 
        \varepsilon     & \varepsilon & \varepsilon \\ 
        1               & 1             & 1
    \end{pmatrix}  
    \sim Y_\ell^T 
    \, .
\end{equation}
Since the largest suppression is $\varepsilon^4$, the minimal $Z_N$ symmetries capable of enforcing such a scenario are $G_f=Z_5$ if SUSY is imposed, and $Z_8$ otherwise.

Since SUSY requires two Higgs doublets with opposite hypercharges~\cite{Martin:1997ns} and the respective Yukawa structure, the ratio between the doublets' vevs ($t_\beta$) becomes relevant.  
Then, $t_\beta^{-1} \sim \varepsilon^2$ can be responsible for the hierarchies between the up- and down-sectors, demanding some of the $Y_{d, \ell}$ interactions to be present at the renormalizable level.    
Finally, the connection of $Y_d \sim Y_\ell^T$ points towards a possible $SU(5)$ Grand Unified Theory embedding, and we show in Table~\ref{tab:charges} a possible $Z_5$ charge assignment compatible with $SU(5)$ multiplet unification. One possible assignment for the non-SUSY variant can be seen in~\cite{Greljo:2024evt}.  

\begin{table}[]
    \centering
    \begin{tabular}{l|c c c c c c }
       field  &  $(Q_1, Q_2, Q_3)$ & $u^c$ & $d^c$ & $L$ & $e^c$ &  $\chi$ \\
 %      $U(1)$ & $\{\omega^3 , \omega^4, 1 \}$ & $\{\omega^3 , \omega^4, 1\}$ & $\omega^3/\omega^4/1$ & $\omega^3/\omega^4/1$ & $\{\omega^3 , \omega^4, 1 \}$ & $\omega^2$ & $\omega$ \\
       $U(1)$ & $(\omega^3 , \omega^4, 1 )$ & $q(Q)$ & $1$ & $q(d^c)$ & $q(Q)$  & $\omega$
    \end{tabular}
    \caption{Possible $Z_5$ (with $\omega^5=1$) charge assignment compatible with a SUSY $SU(5)$ embedding.}
    \label{tab:charges}
\end{table}

%%%%%%%%%%%%%%%%%%%%%%%%%%%%%%%%%%%%%%%%%%%%%%
\section{Domain Walls from $Z_N$ Symmetries}
\label{section: DWs}
%%%%%%%%%%%%%%%%%%%%%%%%%%%%%%%%%%%%%%%%%%%%%%

It is generally considered that gravity does not respect global symmetries, and 
that this explicit breaking through gravity should be small, manifesting itself via Planck-suppressed effective operators. We assume $\mathcal{O}(1)$ coefficients for these operators and no tuning on other parameters and describe (non-)SUSY scalar potentials with $Z_N$ symmetries and the defects that arise.

In non-SUSY theories, 
for $N \leq 4$, the potential can have large barriers between the different minima
\cite{Wu:2022tpe}, with domain wall annihilation leading to potentially observable GW signals. As discussed above, such small $N$ are not viable in the context of minimal (single flavon) dFN models. As we will discuss below, for larger $N$ and in particular for the case of $N=8$ of interest here, the potential has a different shape and the GWs turn out to be unobservable by currently envisioned GW observatories. This is because, for $N \geq 5$, the potential acquires an accidental $U(1)$ symmetry, which is then lifted by a smaller ($Z_N$-symmetric) term, leading to an \textit{orange-squeezer} shape for the potential.

This difference in the shape of the potential can be easily understood by the order at which the Planck-breaking effects and the $U(1)$-breaking ($Z_N$ preserving) operators appear. We assume a naive form for this breaking, as the first non-renormalizable term with a dimensionless coefficient (expected to be $\mathcal{O}(1)$) and suppressed by the Planck scale:
\begin{equation}\label{eq:non susy case}
\begin{split}
    N \leq 4 \, , & \quad \dfrac{V(\chi)}{v_0^4} =  V'_{Z_N} + \dfrac{v_0}{M_P} V'_{\cancel{Z_N}}  \, , \\
    N = 5 \, , & \quad \dfrac{V(\chi)}{v_0^4} = V'_{U(1)}  + \dfrac{v_0}{M_P} V'_{\cancel{Z_N}} +\dfrac{v_0}{\Lambda} V'_{Z_N}  \, , \\
    N \geq 6 \, , & \quad \dfrac{V(\chi)}{v_0^4} = V'_{U(1)}  + \dfrac{v_0}{M_P} V'_{\cancel{Z_N}} +\dfrac{v_0^{N-4}}{\Lambda^{N-4}} V'_{Z_N} \, . 
\end{split}
\end{equation}
where $V'$ is the scalar potential for $\chi/v_0$, and the subscript denotes its symmetry.  
Here, we can see that for $N\leq 4$, the $Z_N$ symmetry is enforced at the renormalizable level, from which we can expect large barriers due to unsuppressed $U(1)$-breaking quartics.
For $N \geq 5$, the $U(1)$ is broken at the same, or at higher order, than the Planck-breaking effects. For the theory to be well-behaved perturbatively, $V'_{Z_N}$ is sub-leading, and thus we expect the potential barriers for DWs to be shallow when compared to the barrier that governs CSs.

When the potential has an \textit{orange-squeezer} shape then both CSs and DWs may form -- or more precisely hybrid defects where the CSs connect to all $Z_N$ domains (see e.g. \cite{Wu:2022tpe, Li:2025gld}. When the domain walls annihilate, the total hybrid defects disappear. Below we therefore focus first on the annihilation of the DWs, and the generated GW signal from this process, and then also conclude that one can estimate that the GWs from the CSs are unobservable as well.

In SUSY models with an R-symmetry (which may contain as a remnant R-parity, protecting against proton decay), the $\chi$ superfield has no R-charge to allow for general Yukawa terms. With an R-charged driving field $S$ we can write a simple superpotential term:
\begin{equation}
    W = S \left( \dfrac{\chi^N}{\Lambda^{N-2}} - M^2 \right) \, .
\end{equation}
The ensuing scalar potential reads
\begin{equation}\label{eq:sym pot}
    V(\chi)\bigg\lvert_{\langle S\rangle} 
    = \left(\dfrac{\lvert \chi \rvert^{2N}}{\Lambda^{2N-4}} - 2 M^2 \dfrac{\text{Re}(\chi^N)}{\Lambda^{N-2}} + M^4 \right) .
\end{equation}
with $\langle S \rangle = 0$ as required by SUSY. 
We define  
\begin{equation}\label{eq:sym vals}
    M^2= v_0^2 \varepsilon^{N-2} \, , \quad \text{with} \quad \varepsilon=\frac{v_0}{\Lambda} \, , 
\end{equation}
with minima $\left<\chi\right>=v_0 e^{i \theta}$, $e^{i N \theta} = 1$. 
Then, 
\begin{equation}    
    V(\chi)\big\lvert_{\langle S\rangle} 
    = v_0^4 \varepsilon^{2N-4}\left( V'_{U(1)}  + V'_{Z_N}  + 1 \right) ,
\end{equation}
where $V'$ is scalar potential for $\chi/v_0$.  
Comparing to non-SUSY cases, in SUSY the $U(1)$-symmetric term 
appears at the same order as the $Z_N$-symmetric term.  
As a consequence, the barriers between the different minima can always be sizeable when compared to $V(0)$, and we can reliably expect DWs as the (dominant) topological defect.

For the DWs to be unstable, we have to lift the degeneracy between the different symmetric vacua.  
In non-SUSY models, it suffices to allow for the $Z_N$ symmetry to be broken, in our case via Planck-suppressed operators - cf. Eq.~\eqref{eq:non susy case} - and barring fine-tuning, the degeneracy of the minima will be lifted.
In the SUSY case, including the leading Planck-suppressed $Z_N$ breaking term in the superpotential shifts the minima but does not lift the degeneracy.  
If SUSY is unbroken, the minima of the potential lie at $V(S, \chi)=0$, 
such that the vevs are the $N$ distinct roots of the F-term, with all $N$ solutions degenerate:
\begin{subequations}
\begin{align}
    &W=S \left( \dfrac{\chi^N}{\Lambda^{N-2}} + \lambda e^{i \alpha} \dfrac{\chi^3}{M_P} - M^2\right) \equiv S f(\chi) \, , \\
    &F_S = f(\langle\chi\rangle) =0 \, ,\qquad F_\chi = \langle S\rangle f'(\langle\chi\rangle) =0 \, , \\
    &V(S, \chi) = \left\lvert F_S\right\rvert^2 + \left\lvert F_\chi\right\rvert^2 \to V(\langle S \rangle, \langle \chi \rangle) =0 \, .
\end{align}
\end{subequations}

Unstable DWs can only arise if both the $Z_N$ symmetry and SUSY are broken \cite{Takahashi:2008mu, Dine:2010eb, Kadota:2015dza}.
We follow a generic approach and introduce the type of SUSY soft-breaking terms that are present in the Minimal SUSY SM (MSSM): squared-masses, and trilinears that respect the structure of the superpotential, with different coefficients. 
 We ignore here additional $U(1)$-symmetric terms that might be relevant for ensuring domain wall formation, e.g.\ K\"ahler corrections inducing a negative mass squared for $\chi$ around zero. The potential then reads ($V\equiv V(S, \chi)$) 
\begin{equation}\label{eq: soft pot}
\begin{aligned}
    V  =& \left\lvert f(\chi) \right\rvert^2 + \lvert S \rvert^2 \left\lvert f'(\chi) \right\rvert^2 + m_S^2 \lvert S \rvert^2 + m_\chi^2 \lvert \chi \rvert^2  \\ 
      + & m_\text{SUSY} \left[ S \left( \mathbf{c}_1  M^2 +  \dfrac{\mathbf{c}_2 \chi^N}{\Lambda^{N-2}} +  \dfrac{\mathbf{c}_3 \chi^3}{M_P} \right) + \text{h.c.} \right]  \\
    \equiv & \lvert f(\chi) \rvert^2 + m_{S, \text{eff}}^2 \lvert S \rvert^2 + m_\text{SUSY} \big[ S N(\chi) + \text{h.c.}\big] \, ,
\end{aligned}
\end{equation}
with $\mathbf{c}_i \equiv c_i e^{i \gamma_i}$, and 
\begin{subequations}
\label{eq: defs}
\begin{align}    
    m^2_{S, \text{eff}} &= m_S^2 + \lvert f'(\chi)\rvert^2 \, , \\
    N(\chi) &= \left(  \mathbf{c}_1 M^2 +  \mathbf{c}_2 \dfrac{\chi^N}{\Lambda^{N-2}} + \mathbf{c}_3 \dfrac{\chi^3}{M_P} \right) \, .
\end{align}
\end{subequations}

As a consequence $S$ gets a non-zero vev:
\begin{equation}
    \dfrac{\partial V}{\partial S} =0     \Leftrightarrow \langle S \rangle = -m_\text{SUSY}\dfrac{N^*(\chi^\dagger)}{m^2_{S, \text{eff}}} \, .
\end{equation}
We obtain $V$ with $S$ at its minimum: 
\begin{equation}\label{eq:pot chi}
    V(\chi)\bigg\rvert_{\langle S \rangle} = \lvert f(\chi)\rvert^2 +m_\chi^2 \lvert \chi\rvert^2 - m^2_\text{SUSY}\dfrac{\lvert N(\chi)\rvert^2}{m^2_{S, \text{eff}}} \, , 
\end{equation}
where we note that $m^2_{S, \text{eff}}$ depends on $\chi$, $\chi^\dagger$.  
We restrict ourselves in the following to the case $m_\chi^2<0$, to ensure that $\chi=0$ is a local maximum of $V(\chi)$.

%%%%%%%%%%%%%%%%%%%%%%%%%%%%%%%%%%%%%%%%%%%%%%
\section{Gravitational Wave Signal: General Considerations}
\label{section: GWs}
%%%%%%%%%%%%%%%%%%%%%%%%%%%%%%%%%%%%%%%%%%%%%%

The GW spectrum is controlled by the peak frequency ($f_p$) and amplitude ($\Omega_p$) as~\cite{Notari:2025kqq}
\begin{equation}
    \Omega_\text{gw}(f) = \Omega_p \, \mathcal{S}(f) \, ,
\end{equation}
where $\mathcal{S}(f)$ is the spectral shape function 
\begin{equation}
    \mathcal{S}\left(f\right) = \dfrac{3+\beta+(f_p/f_b)^{\beta+\gamma}}{\beta\left(\dfrac{f}{f_p}\right)^{-3}+3\left(\dfrac{f}{f_p}\right)^\beta+\left(\dfrac{f_b}{f_p}\right)^{-\beta}\left(\dfrac{f}{f_b}\right)^\gamma} \, , 
\end{equation}
with $\beta\simeq 1/2$, $\gamma\simeq 1.8$, and $f_b\simeq 2.8 f_p$ as given in~\cite{Notari:2025kqq}.

When the DWs annihilate during the radiation era, $f_p$ and $\Omega_p$ are computed via~\cite{Saikawa:2017hiv}
\begin{subequations}\label{eq:Peak Defs}
    \begin{align}
    \dfrac{f_p}{10^{-9} \text{ Hz }} =& 1.1 \left(\dfrac{g_*(T_\text{ann})}{10} \right)^{\frac{1}{6}} \left( \dfrac{T_\text{ann}}{10^{-2} \text{ GeV}} \right) \, ,\\
    \dfrac{\Omega_p h^2}{7.2 \times 10^{-18}} =&  \tilde{\epsilon}_\text{gw} \mathcal{A}^2  \left(\dfrac{g_{*}(T_\text{ann})}{10} \right)^{-\frac{4}{3}} \times \nonumber \\
    &\times \left( \dfrac{\sigma}{\text{TeV}^3} \right)^2 \left( \dfrac{T_\text{ann}}{10^{-2} \text{ GeV}} \right)^{-4} \, , \label{eq:peakOmega}
\end{align}
\end{subequations}
where $\sigma$ is the domain wall tension, $T_\text{ann}$ is the temperature at which the DWs annihilate, $\mathcal{A}$ is called an area parameter, $\tilde{\epsilon}_\text{gw}$ is the efficiency parameter (cf.\  \cite{Hiramatsu:2012sc, Hiramatsu:2013qaa} for values), and where we have taken the approximation that the radiation and entropy relativistic degrees of freedom ($g_{*}$ and $g_{*s}$) are identical.  
The annihilation temperature is given by~\cite{Saikawa:2017hiv}
\begin{equation}\label{eq:Tann Def}
\begin{aligned}
      T_\text{ann}  &= 3.41 \times 10^{-2} \text{ GeV}  (C_\text{ann} \, \mathcal{A})^{-\frac{1}{2}} \\
      & \times \left( \dfrac{g_*(T_\text{ann})}{10} \right)^{-\frac{1}{4}} 
      \left( \dfrac{\sigma}{\text{TeV}^3} \right)^{-\frac{1}{2}} \left( \dfrac{\Delta V}{\text{MeV}^4}\right)^{\frac{1}{2}} \, , 
\end{aligned}
\end{equation}
with $C_\text{ann}$ being an $\mathcal{O}(1)$ coefficient which we take from ref.~\cite{Kawasaki:2014sqa} for the $N=5$ case when we present numerical results, and with $\Delta V$ being the difference of $V(\chi)$ at the two minima.

The GW spectrum will be primarily controlled by two relevant quantities: the wall tension, $\sigma$, and the bias between two minima, $\Delta V$. As such, we focus now on these quantities for the (non-)SUSY cases.

\subsection{The non-SUSY case}

We start by looking at the wall tension for a $Z_N$ symmetric non-SUSY theory.  
We neglect the Planck-breaking effects, as these should be small.
We can describe the $Z_N$ symmetric potential as ($\chi= \rho\,  e^{i\theta}$):
\begin{equation}
    \begin{aligned}
         &V_{Z_N}  = \lambda_0 \left(\rho^2 - v_0^2 \right)^2 +& \\
      &\dfrac{\lambda_N}{\Lambda^{N-4}} \bigg[ -2 \rho^N \cos\left(N \theta\right) + N v_0^{N-2} \rho^2 - (N-2)v_0^N \bigg]&     \, ,
    \end{aligned}
\end{equation}
where we have uplifted the symmetric potential $V_{Z_N}$ such that $V_{Z_N}(v_0) =0$, and ignored irrelevant $U(1)$-symmetric corrections to the renormalizable term.  
We note that the barrier height between different minima is shallow when compared to the barrier at $\chi=0$:
\begin{subequations}
    \begin{align}
        V(0) &=  v_0^4 \bigg[ \lambda_0  - (N-2) \lambda_N \varepsilon^{N-4} \bigg] \, , \\
        V(\chi_\text{b}) &= 4 v_0^4 \lambda_N \varepsilon^{N-4} \, , \quad \text{with} \quad \chi_\text{b} = v_0 e^{i \frac{2 \pi (k+1)}{N}} \, , 
    \end{align}
\end{subequations}
so the \textit{orange-squeezer} shape of the scalar potential should make it more viable to cross between minima through the barriers, rather than the peak at $\chi=0$.  
We can approximate the transition between adjacent minima by taking into account the axial mode only: 
\begin{align}
    V\left(v_0 e^{i \theta}\right) =  2 \lambda_N \varepsilon^{N-4} v_0^4 \bigg[ 1 - \cos\left(N \theta \right) \bigg] \, .
\end{align}
We can take advantage of the similarities and relate to the potentials that usually arise in the context of axion models, which allows us to approximate the wall tension as (assuming adjacent minima)~\cite{Saikawa:2017hiv}:
\begin{subequations}\label{eq:axionpot}
    \begin{align}
        V(a) =  \dfrac{m^2 v^2}{N^2} \left[ 1 - \cos\left(N \dfrac{a}{v} \right) \right], \quad
        \sigma = \dfrac{8 m v^2}{N^2} \, , 
    \end{align}
\end{subequations}
which in our case becomes 
\begin{equation}\label{eq:sigmanonSUSY}
    \sigma =  \dfrac{8 \sqrt{2} \sqrt{\lambda_N } \varepsilon^\frac{N-4}{2} v_0^3}{N} \, . 
\end{equation}
$\Delta V$ is governed by the Planck-breaking terms (with $\lambda_j e^{i \alpha_j}$ being the coefficients):
\begin{equation}
    \begin{aligned}
        V_{\cancel{Z_N}}  = \dfrac{2 \rho^5}{M_P}  \bigg[ & \lambda_1 \cos\left(\alpha_1 + 5 \theta \right) + \lambda_2 \cos\left(\alpha_2 + 3 \theta \right) \\
        &+ \lambda_3 \cos\left(\alpha_3 + \theta \right) \bigg] \, . 
    \end{aligned}
\end{equation}
If the phases of each term do not conspire to cancel each contribution of $V_{\cancel{Z_N}}$ to an arbitrary extent, the lifting of the degeneracy of the minima is approximately: 
\begin{align}
    \Delta V \sim  \dfrac{v_0^5}{M_P} \, .
\end{align}

Assuming the dFN motivation, implying $\mathcal{O}(1)$ coefficients, setting $\varepsilon=(0.19)^2$ (a best-fit representative from a dFN analysis) and with $N=8$, we get a GW spectrum with unobservably low amplitude in the frequency region of interest.  
If we wish to have observable GW imprints under these circumstances, the Planck-breaking terms would have to conspire 
to provide a cancellation of 7 to 8 orders of magnitude (or have another way to suppress these contributions, such as non-perturbative gravity). 
Furthermore, for peak frequencies between $10^{-9}$ and $10^2$, $v_0$ is predicted in the range from $10^{-2}$ GeV to $10^9$ GeV, from where we estimate that the GW signal from the CS component of the hybrid defect is unobservable as well (see e.g. \cite{Blanco-Pillado:2024aca}).

\subsection{The SUSY case}

To find the DW tension, we neglect the SUSY-breaking and Planck-breaking sectors. 
Obtaining the field profiles in the symmetric limit, defined by Eqs.~\eqref{eq:sym pot}~and~\eqref{eq:sym vals}, computing the relevant field profile and integrating the energy density~\cite{Saikawa:2017hiv, Wu:2022stu}, we find 
\begin{equation}\label{eq:sigmaSUSY}
    \sigma = \mathcal{O}(1)\, v_0^3 \varepsilon^{N-2} \, ,
\end{equation}
where the $\mathcal{O}(1)$ coefficient depends on whether the two minima are adjacent.

Estimating $\Delta V$, i.e.\ the lifting of the degeneracy, requires including both Planck- and SUSY-breaking terms. From Eqs.~\eqref{eq:pot chi}~and~\eqref{eq: defs},
and requiring small Planck-breaking terms compared to the symmetry-allowed terms, we expect small deformations of the symmetric potential.
We estimate $\Delta V$ by assuming the shifts of the minima to be negligible (recall the discussion around $m_\chi^2 \lesssim 0$).  
Then, $\Delta V$ should be governed by the explicit $Z_N$ breaking terms in $V(\chi)$, where the suppression is explicit, and by the small parameter $y$ (note that this parameter must necessarily be small for the Planck-breaking term to be subdominant with respect to the $Z_N$ symmetric terms):
\begin{equation}\label{eq:y}
    y = \dfrac{v_0}{M_P \varepsilon^{N-2}} \, .
\end{equation}
Expanding $V(\chi)$ in $y$, and taking $m_{S} \ll v_0 \varepsilon^{N-2}$, we find: 
\begin{align}
     \Delta V \approx &\dfrac{v_0^2 m_\text{SUSY}^2}{N^2} \left(\dfrac{v_0}{M_P \varepsilon^{N-2}} \right) \nonumber \\
    \bigg\{ &\left\lvert \mathbf{c}_1 +\mathbf{c}_2  \right\rvert^2 \dfrac{6 \lambda }{N} \Delta \cos\left(2\beta+\alpha\right)  \nonumber \\ 
    &  - 2 c_1 c_3 \Delta\cos\left(3\beta+\gamma_3-\gamma_1\right) \nonumber \\
    & - 2 c_2 c_3 \Delta\cos\left(3\beta+\gamma_3-\gamma_2\right) \bigg\}  \, , \label{eq:deltaPot2}
\end{align}
where $\Delta f(x) = f(x_1) - f(x_2)$.  
Eq.~\eqref{eq:deltaPot2} shows the dependency on the flavour symmetry breaking scale and on the SUSY scale.
SUSY dFN models can have enhanced GW signals compared to their non-SUSY counterparts when $m_\text{SUSY}^\text{eff} \ll v_0$, effectively suppressing $\Delta V$ without affecting $\sigma$.  

%%%%%%%%%%%%%%%%%%%%%%%%%%%%%%%%%%%%%%%%%%%%%%
\section{Results}
\label{section: GWs - 2}
%%%%%%%%%%%%%%%%%%%%%%%%%%%%%%%%%%%%%%%%%%%%%%

\begin{figure}
    %\centering
    \includegraphics[width=1\linewidth]{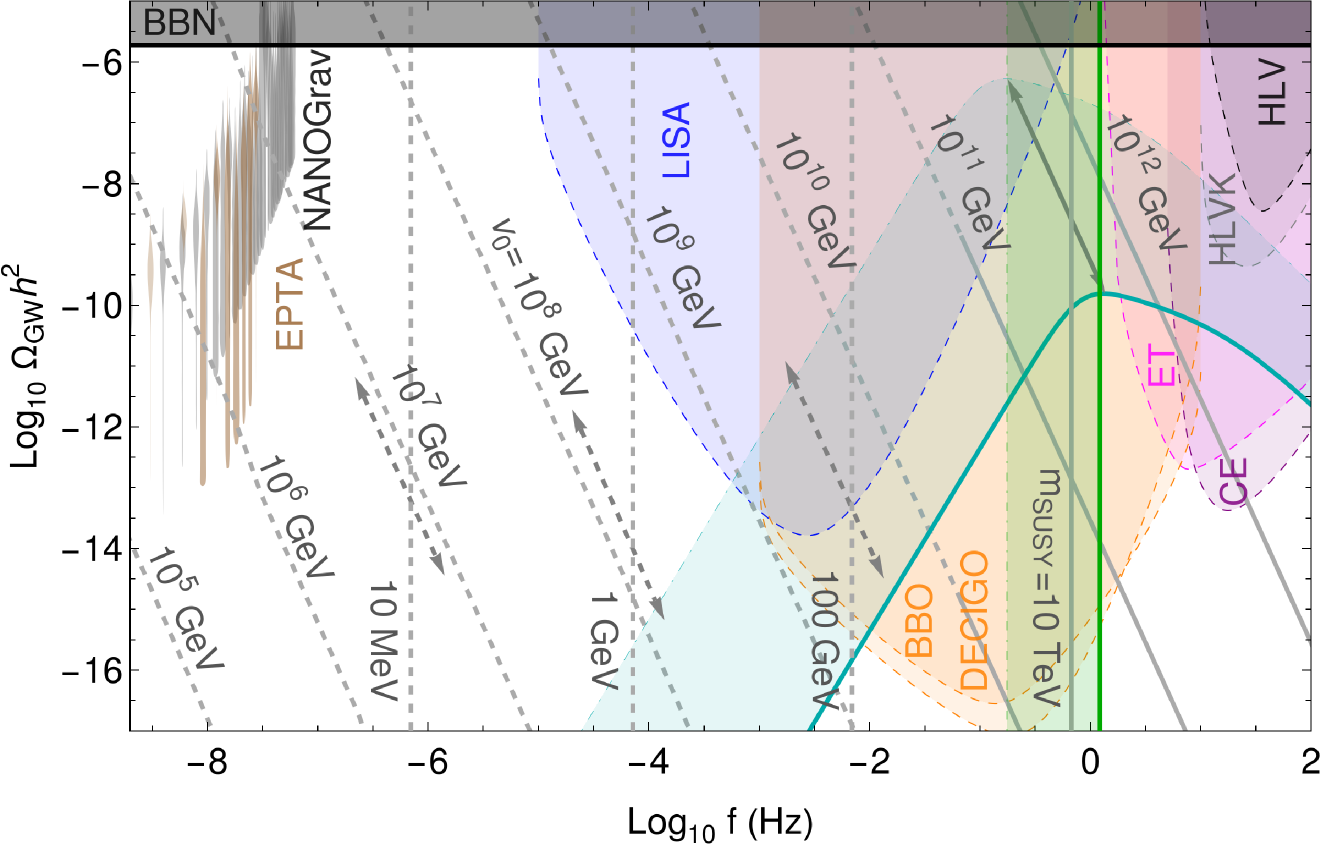}
    \caption{
    In teal, the GW spectrum for a benchmark case with $m_\text{SUSY} = 10$ TeV and $v_0 \approx 5 \times  10^{11}$ GeV.
    In gray (dashed for $m_\text{SUSY} < 1$ TeV), peak amplitude and frequency for $v_0$ and $m_\text{SUSY}$ isocurves.    
    The arrows (gray) and band (green) show the 90\% shift of the peak positions due to $\mathcal{O}(1)$ coefficients.  
    Experimental sensitivities are taken from~\cite{Schmitz:2020syl} and for details on the nucleosynthesis (BBN) bound see e.g.\ \cite{Maggiore:1999vm}. 
    \label{fig:PLOT} }
\end{figure}

A simple estimate shows that the non-SUSY dFN scenario leads to unobservable GW signatures. We therefore focus here solely on the SUSY scenario and fix $N=5$ and $\varepsilon = (0.19)^2$ to relate to the dFN results.  
Since we are interested in the general behaviour with $v_0$ and $m_\text{SUSY}$, we compute $(f_p, \Omega_p h^2)$ for either fixed values of $m_\text{SUSY}$ or $v_0$.    
The coefficients are scanned over ranges:  
$\lambda \in [0.1,1]$; 
$c_i$, $m_S^2$ and $-m_\chi^2$ $\in [.3,3]$ 
(such that the SUSY masses average around $m_\text{SUSY}$);  
and all phases are scanned over their domain.  
The result is shown in Figure~\ref{fig:PLOT}. 
In gray, we show the $(f_p, \Omega_p h^2)$ isocurves for $v_0$ and $m_\text{SUSY}$, with each isocurve labelled.  
The arrows encapsulate the effect of varying the coefficients and phases, showing a 90\% band where the scanned points lie.  
The isocurves are chosen specifically such that they fall on the median frequency of the band. 
The asymmetry of the arrow with respect to the line is expected since the phases can be fine-tuned to provide an arbitrarily small (but not arbitrarily large) $\Delta V$, effectively pushing the GW signal arbitrarily up. 
For $m_\text{SUSY}=10$ TeV we show the frequency band explicitly (in green), and provide the GW spectrum for a benchmark point (in teal) that would lie comfortably within reach of future experiments.

The $m_\text{SUSY}$ isocurves are vertical, meaning $m_\text{SUSY}$ defines the frequency of the GW peak, \textit{i.e.}\, $\Delta V/ \sigma$ is independent of $v_0$ (cf. Eqs.~\eqref{eq:Peak Defs}~and~\eqref{eq:Tann Def}), agreeing with the expectation following Eq.~\eqref{eq:deltaPot2}.  
Deviation from this behaviour could signal a regime where the approximations of the last section do not hold, but we do not see this within the range of Fig.~\ref{fig:PLOT}.  

Since we compute the DW tension in the symmetric limit, $\sigma$ is determined by $v_0$ alone.  
Then, the precise values of any of the other parameters we scan over
only affect $\Delta V$. Thus, the frequency is fixed by $\sqrt{\Delta V/v_0^3} \propto m_\text{SUSY}$.  
Any change in these parameters
are equivalent to a change in $m_\text{SUSY}$ for a different set of parameters, and so we understand why the slope of the arrows in Fig.~\ref{fig:PLOT} must necessarily be similar to the $v_0$ isocurves: it encapsulates the effect of changing $\Delta V$ (\textit{i.e.,} $m_\text{SUSY}$).

Taking the expectation that $m_\text{SUSY}$ should be around the same scale as the MSSM sparticle spectrum (which we assume to compute $g_*(T_\text{ann}$), it is clear that explaining the 2023 Pulsar Timing Array signal \cite{NANOGrav:2023gor,EPTA:2023fyk, Reardon:2023gzh,Xu:2023wog} is incompatible with collider data~\cite{ParticleDataGroup:2024cfk}. 
Indeed, any GW signal peak below $10^{-1}$ Hz should be hard to accommodate given the LHC limits from SUSY searches.  
A possible workaround could come \textit{e.g.}\ in the form of non-perturbative gravity leading to an instanton enhancement of the effective $M_P$.

The observed value of the Higgs mass at 125 GeV points towards the SUSY spectrum (most notably the stops) to lie around the $10$ TeV range~\cite{Giudice:2011cg}.  
Interestingly, we see that this expectation selects a frequency range at a golden spot for future experiments, with the GW signal in a frequency range (cf.\ green band in Fig.~\ref{fig:PLOT}) potentially detectable by
LISA~\cite{LISA:2017pwj},
DECIGO~\cite{Seto:2001qf},
BBO~\cite{Corbin:2005ny},
Einstein Telescope~\cite{Sathyaprakash:2012jk} and
Cosmic Explorer~\cite{LIGOScientific:2016wof} simultaneously, 
if $v_0$ lies between $10^{11}$ and $10^{12}$ GeV.  

Note that regardless if a GW signal is measured in the future, we can make statements on SUSY $Z_N$ models with a single flavon.  
The condition that the symmetry holds approximately (sub-dominant Planck-Breaking terms) places an upper bound (cf. Eq.~\eqref{eq:y}):
\begin{equation}
     \varepsilon^{N-2} \gg \dfrac{v_0}{M_P} \, .
     \label{eq:upper_gen}
\end{equation}
Moreover, linking the $Z_N$ to a dFN model imposes a stricter constraint, from the requirement that the Planck-suppressed contribution does not exceed the smallest entry of the Yukawa matrices allowed by the $Z_N$ group:
\begin{equation}
     \varepsilon^{N-1} \gtrsim \dfrac{v_0}{M_P} \, \Rightarrow v_0 \lesssim 10^{13} \text{ GeV} \, .
     \label{eq:upper_dFN}
\end{equation}
In Fig.~\ref{fig:PLOT}, note the Hanford (LIGO), Livingston (LIGO), and Virgo (HLV) \cite{LIGOScientific:2014pky, VIRGO:2014yos} region is close to this bound, meaning a signal detected by this experiment would have been at the edge of the validity region of our scenario.

Finally, we note that regardless of any observation of GW signals, assuming that the cosmological evolution is such that the DWs form and are not diluted afterwards by a phase of inflation, the nucleosynthesis bound~\cite{Maggiore:1999vm} can place an upper bound on the flavour breaking scale, if this symmetry breaking is post-inflationary and DWs form.  
For the specific case of $m_\text{SUSY} = 10$ TeV, we see that the $Z_N$ breaking scale must be $v_0 \lesssim 2\times 10^{12}$ GeV. More importantly, the non-observation of GW signals by future experiments would lower this to $v_0 \lesssim 5 \times 10^{10}$ GeV, whereas complimentary searches by colliders (\textit{e.g.}\ through flavour-changing neutral currents) can provide lower bounds, potentially probing the model in the future.

%%%%%%%%%%%%%%%%%%%%%%%%%%%%%%%%%%%%%%%%%%%%%%
\section{Conclusions}
\label{section: conclusion}
%%%%%%%%%%%%%%%%%%%%%%%%%%%%%%%%%%%%%%%%%%%%%%

We have investigated stochastic gravitational-wave background signals from unstable domain walls in the context of models that aim to explain the observed charged-fermion mass hierarchies through flavour symmetries. In particular, we focused on minimal supersymmetric and non-supersymmetric realisations involving a single flavon field and discrete $Z_N$ symmetries with the smallest viable value of $N$, subject to our phenomenological requirements. DWs are unstable due to the presence of $Z_N$-violating Planck-suppressed effective operators. We found that the minimal non-supersymmetric models based on a $Z_8$ symmetry do not yield observable gravitational-wave signals. By contrast, domain-wall annihilation in the minimal supersymmetric $Z_5$ scenario gives rise to a stochastic gravitational-wave background whose spectrum, for supersymmetric mass scales $m_\text{SUSY}$ in the multi-TeV range, falls within the frequency windows of LISA, DECIGO, BBO, Einstein Telescope, and Cosmic Explorer. In addition to $m_\text{SUSY}$, the peak amplitude is governed by the flavour-hierarchy parameter $\varepsilon$, as determined from a fit to the charged-fermion masses, and the flavour-symmetry breaking scale $v_0$. Current constraints from DWs for $m_\text{SUSY}=10$ TeV imply an upper bound of $v_0 \lesssim 2\times 10^{12}\,\mathrm{GeV}$, which could strengthen to $v_0 \lesssim 5 \times 10^{10}\,\mathrm{GeV}$ in the absence of a signal at future experiments (assuming DWs form and are not diluted away after their formation). The discovery of a peaked gravitational-wave spectrum in the relevant frequency range, interpreted within the class of models considered here, could offer remarkable insight into the origin of flavour.

%%%%%%%%%%%%%%%%%%%%%%%%%%%%%%%%%%%%%%%%%%%%%%
\section*{Acknowledgements}
%%%%%%%%%%%%%%%%%%%%%%%%%%%%%%%%%%%%%%%%%%%%%%

IdMV thanks the University of Basel for hospitality.
IdMV acknowledges funding from Fundação para a Ciência e a Tecnologia (FCT) through the FCT Mobility program, and through
the project CFTP-FCT Unit 
UID/00777/2025 (\url{https://doi.org/10.54499/UID/00777/2025}).

\bibliographystyle{utphys2}

\end{document}